\documentclass[lettersize,journal]{IEEEtran}
\def\BibTeX{{\rm B\kern-.05em{\sc i\kern-.025em b}\kern-.08em
    T\kern-.1667em\lower.7ex\hbox{E}\kern-.125emX}}
\usepackage{textcomp}
\usepackage{svg}
\usepackage{epsfig}
\usepackage{epstopdf}
\usepackage{amsmath}
\usepackage{amssymb}
\usepackage{amsfonts}
\usepackage{algorithmic}
\usepackage{adjustbox}
\usepackage{array}
\usepackage{booktabs}
\usepackage{hyperref}
\usepackage{cleveref}
\usepackage{graphicx} 
\usepackage{siunitx}
\usepackage{float}
\usepackage{multirow}
\usepackage{cite}
\usepackage{bm}
\usepackage{color}
\usepackage{subcaption}
\usepackage[normalem]{ulem}
\usepackage{balance}
\newcolumntype{M}[1]{>{\centering\arraybackslash}m{#1}}
\usepackage{tabularx,ragged2e}

\usepackage{scalerel}
\usepackage{tikz}
\usetikzlibrary{svg.path}
\definecolor{orcidlogocol}{HTML}{A6CE39}
\tikzset{
	orcidlogo/.pic={
		\fill[orcidlogocol] svg{M256,128c0,70.7-57.3,128-128,128C57.3,256,0,198.7,0,128C0,57.3,57.3,0,128,0C198.7,0,256,57.3,256,128z};
		\fill[white] svg{M86.3,186.2H70.9V79.1h15.4v48.4V186.2z}
		svg{M108.9,79.1h41.6c39.6,0,57,28.3,57,53.6c0,27.5-21.5,53.6-56.8,53.6h-41.8V79.1z M124.3,172.4h24.5c34.9,0,42.9-26.5,42.9-39.7c0-21.5-13.7-39.7-43.7-39.7h-23.7V172.4z}
		svg{M88.7,56.8c0,5.5-4.5,10.1-10.1,10.1c-5.6,0-10.1-4.6-10.1-10.1c0-5.6,4.5-10.1,10.1-10.1C84.2,46.7,88.7,51.3,88.7,56.8z};
	}
}
\newcommand\orcidicon[1]{\href{https://orcid.org/#1}{\mbox{\scalerel*{
				\begin{tikzpicture}[yscale=-1,transform shape]
				\pic{orcidlogo};
				\end{tikzpicture}
			}{|}}}}
\title{\LARGE \bf
Breast Lump Detection and Localization \\with a Tactile Glove Using Deep Learning}
\author{Togzhan Syrymova$^{\orcidicon{0000-0003-1032-5880}}$, Amir Yelenov$^{\orcidicon{0000-0003-0674-5460}}$ , Karina Burunchina$^{\orcidicon{0009-0005-9391-3013}}$, Nazgul Abulkhanova$^{\orcidicon{0000-0003-3526-5159}}$, \\Huseyin Atakan Varol$^{\orcidicon{0000-0002-4042-425X}}$,~\IEEEmembership{Senior Member,~IEEE,} Juan Antonio Corrales Ram\'on$^{\orcidicon{0000-0002-9373-7954}}$, \\Zhanat Kappassov$^{\orcidicon{0000-0003-3262-3993}}$,~\IEEEmembership{Senior Member,~IEEE}
\thanks{This work was funded by MSHE Kazakhstan Grant number AP23485994,  by Nazarbayev University under FDCRGP no. 201223FD2606}
\thanks{T. Syrymova is with the Dept. of Robotics Engineering, Nazarbayev University , 53 Kabanbay Batyr Ave, Astana, 010000, Kazakhstan and 
Centro Singular de Investigaci\'on en Tecnolox\'ias Intelixentes~(CiTIUS), Universidade de Santiago de Compostela, 15782 Santiago de Compostela, Spain (email: togzhan.syrymova@nu.edu.kz).}
\thanks{J.A.Corrales Ram\'on is with Centro Singular de Investigaci\'on en Tecnolox\'ias Intelixentes (CiTIUS), Universidade de Santiago de Compostela, 15782 Santiago de Compostela, Spain (email: jacr56@gmail.com)}
\thanks{A. Yelenov, and Z. Kappassov are with the Dept. of Robotics Engineering, Nazarbayev University, 53 Kabanbay Batyr Ave, Astana, 010000, Kazakhstan (email: amir.yelenov@nu.edu.kz, zhkappassov@nu.edu.kz). }
\thanks{N. Abulkhanova is with the University Medical Center Corporate Fund~(UMC), 32 Turan Ave, Astana, 010000, Kazakhstan. (email: 7571768@gmail.com)}
\thanks{H.A. Varol and K. Burunchina are with the Institute of Smart Systems and Artificial Intelligence, Nazarbayev University, Astana, 010000, Kazakhstan (email: ahvarol@nu.edu.kz, foxake1610@gmail.com)}

\thanks{Corresponding author: Z. Kappassov}}
\date{}
\begin{document}
\maketitle
\begin{abstract}
Breast cancer is the leading cause of mortality among women. Inspection of breasts by palpation is the key to early detection. We aim to create a wearable tactile glove that could localize the lump in breasts using deep learning~(DL).
In this work, we present our flexible fabric-based and soft wearable tactile glove for detecting the lumps within custom-made silicone breast prototypes~(SBPs).  SBPs are made of soft silicone that imitates the human skin and the inner part of the breast. Ball-shaped silicone tumors of 1.5-, 1.75- and 2.0-cm diameters are embedded inside to create another set with lumps. Our approach is based on the InceptionTime DL architecture with transfer learning between experienced and non-experienced users. We collected a dataset from 10 naive participants and one oncologist-mammologist palpating SBPs. We demonstrated that the DL model can classify lump presence, size and location with an accuracy of 82.22\%, 67.08\% and 62.63\%, respectively. 
In addition, we showed that the model adapted to unseen experienced users with an accuracy of 95.01\%, 88.54\% and 82.98\% for lump presence, size and location classification, respectively. This technology can assist inexperienced users or healthcare providers, thus facilitating more frequent routine checks. 
 
\end{abstract}
\begin{IEEEkeywords}
Tactile sensing, lump detection, tactile glove, silicone breast prototypes, deep learning.
\end{IEEEkeywords}
\section{Introduction}
\begin{figure}[!t]
    \centering
    \includegraphics[width = .8\linewidth]{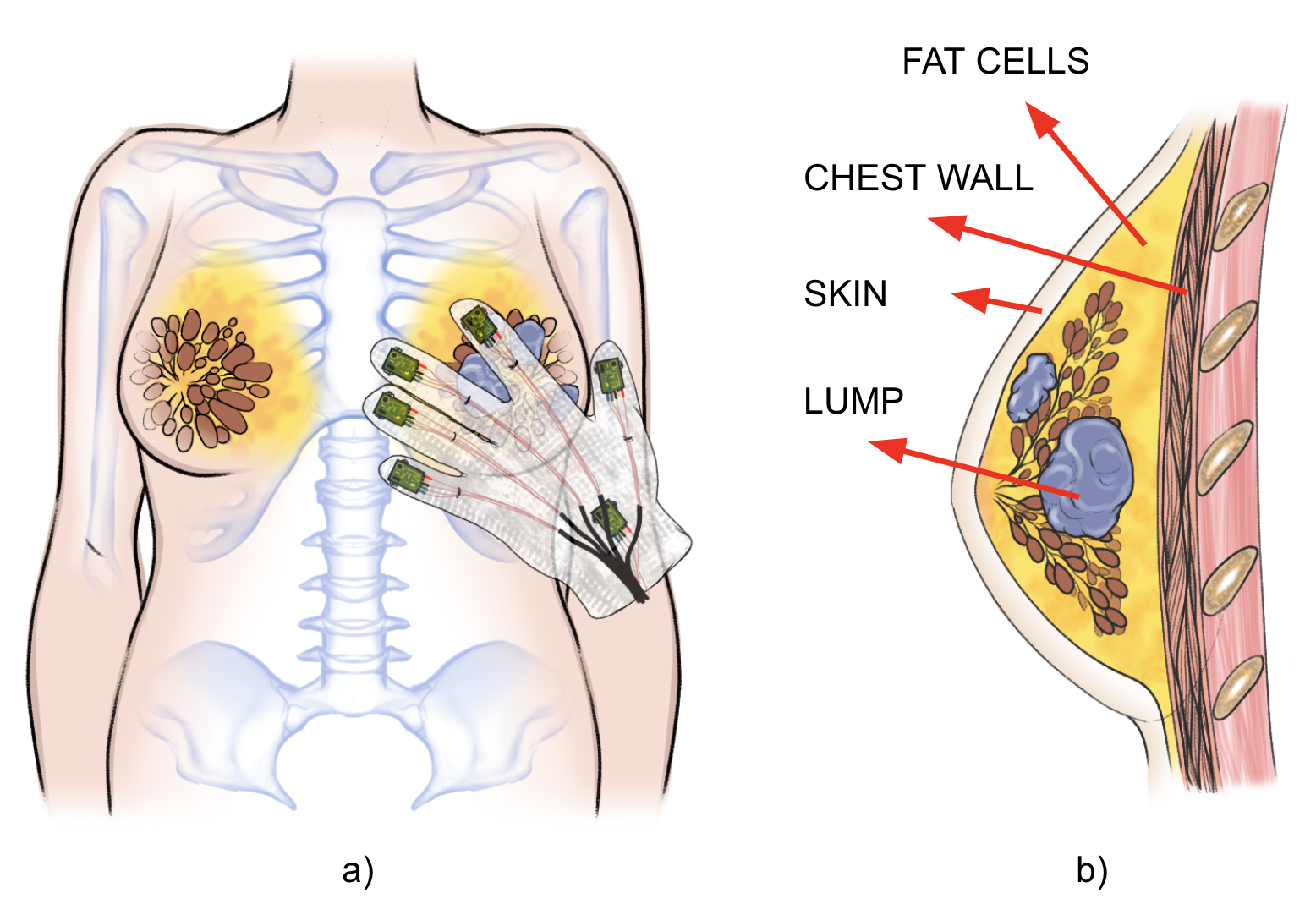}
    \caption{a) The experimental scenario of the Tactile Glove for palpation of the breast with a lump, and b) side view of the breast structure with a lump.}
    \label{fig:breast_with_nodule}
\end{figure}

\IEEEPARstart{W}{ith} 1.98 million cases and over half a million deaths, breast cancer was reported as the leading cause of mortality among women in 2015~\cite{global_burden_of_disease_2019_cancer_collaboration_cancer_2022}.
Early diagnosis by means of screening can reduce cancer mortality and morbidity rates, and increase the chances of effective therapy and survival. The initial phase of breast cancer is an early-stage non-metastatic oncology disorder with high chances~(about $70-80\%$) for cure using multimodal therapy~\cite{harbeck_breast_2019}. Considering that breast cancer manifests itself as an abnormal mass in lobules and ducts of the breast~(see Fig.~\ref{fig:breast_with_nodule}), it is vital to detect breast lumps in the preliminary stages. Non-invasive methods for breast pathology examination include imaging-based~(mammography, magnetic resonance imaging (MRI), and ultrasound) and palpation-based approaches~(breast self-examination~(BSE) and clinical breast examination~(CBE))~\cite{bonadonna_hortobagyi_valagussa_2006}. 
    
\par Mammography is one of the commonly used techniques not only for diagnosis but also for screening. The procedure involves the compression of the breast tissue between two plates and the application of low doses of X-ray radiation to image the contrast between the lump and surrounding tissues. Weigel et al.~\cite{weigel_digital_2017} conducted a study with 25,729 women and reported the sensitivity of digital mammography as $79.9\%$. The study revealed a significant degradation in the diagnostic performance for women with denser breasts. Therefore, women with dense and vascular breasts, e.g., younger women~\cite{weigel_digital_2017}, can benefit from supplementary screening~\cite{melnikow_supplemental_2016}. Moreover, mammography is not recommended for women with risk-increasing genetic variations and/or hereditary breast cancer proclivity since X-ray radiation elevates the risk of breast cancer~\cite{colin_radiation_2017}.
    
Another imaging-based method, ultrasound imaging is used complementarily for the assessment of mammographically or clinically detected masses as it has higher sensitivity in detecting breast cancer in women with denser breasts~\cite{nothacker_early_2009}. Ultrasound imaging is effective but is usually conducted by a trained specialist after a preliminary diagnosis of the lump presence and its approximate location. MRI employs strong magnetic fields to produce a 3D image of the breast. It is a minimally invasive~(a contrast agent is injected into a vein in the arm) and an effective technique for examining dense breasts. However, it is a costly and time-consuming procedure.
    
An abnormal lump that is substantially stiffer than the normal fat tissue is one of the diagnostic markers of breast cancer~\cite{ramiao_biomechanical_2016}. It can be detected by means of sequential palpation using the fingers while gradually increasing the applied force to examine the deeper tissues. Palpation is an easy and painless method that can be performed by a patient themselves~(BSE) or a health care provider~(CBE). 

\par Annual breast cancer screening is recommended for women above the age of 40 years~\cite{smith_american_2003}. Younger women who do not have a predisposition for breast cancer do not undergo screening unless they have specific symptoms such as localized pain, nipple symptoms~(e.g., bleeding), or visual changes~(e.g., skin lesions). 
According to Global Burden of Disease Cancer Collaboration~\cite{cancer_1995_2015}, breast cancer was one of the most common types of cancer among young people~(aged 15-39 years) globally in 2015. 
In some cases, early breast cancer does not exhibit any of the above-mentioned symptoms. Thus, BSE can serve as a primary and cost-effective screening method for young women. However, the efficacy of this method remains limited due to a lack of awareness of the procedure and knowledge of the early signs of breast cancer~\cite{suh_breast_2012, rahman_awareness_2019}.

\par Considering the shortcomings of BSE, CBE is performed by a trained healthcare provider according to a specific protocol. During CBE, a health care provider inspects the breast visually and physically by palpating it to detect abnormalities within.
However, diagnostic accuracy depends highly on the proficiency and training of the healthcare provider. The noticeable difference in the sensitivity of CBE in clinical practice~($28-36\%$) and Canadian National Breast Screening Study~($63\%$) was attributed to the thorough training of the health care providers in the latter prior to the study~\cite{provencher_is_2016}.
    
Presumably, a technology with enhanced sensitivity that could support the healthcare provider during the examination can result in higher preliminary screening efficiency via palpation.
Indeed, during the last decades, various methods were developed for detecting and localizing the abnormalities within the breast~\cite{nguyen_tactile_2014, xu_breast_2013,xu_development_2016,  naidu_low_cost_2017, li_mech_im_tact, xie_fiber, ayyildiz_optoelec, gwilliam_soft_tissue, lee_lesion_char, jia_lump_2013, sahu_tactile, egorov_mi_breast, yildiz_novel, murayama_development_2008}.
Tactile imaging~(TI) is one of the emerging approaches in this research field. It is a non-invasive technique that involves the application of compression on the surface to measure the pressure/force distribution over the contact area. 

The majority of the tactile technologies for breast examination were based on single or array-type resistive~\cite{xu_breast_2013, naidu_low_cost_2017}, piezoelectric~\cite{xu_development_2016, naidu_low_cost_2017, murayama_development_2008} and capacitive~\cite{li_mech_im_tact, gwilliam_soft_tissue} sensors, and camera-based tactile systems~\cite{nguyen_tactile_2014, xie_fiber, ayyildiz_optoelec, sahu_tactile, jia_lump_2013}.
Jia et al.~\cite{jia_lump_2013} employed a Gelsight sensor for lump detection using a support vector machine classifier by pressing the top of the silicone prototype against an elastomeric pad covered with a reflective membrane. Like Jia et al., other researchers also developed optical tactile systems based on compression of the silicone prototypes with the sensor attached at the tip of a vertically moving plate~\cite{ayyildiz_optoelec, lee_lesion_char}. 
Alternatively, Gwilliam et al. utilized a capacitive array sensor to press on a set of cylindrical silicone rubbers with embedded balls of different sizes and depths~\cite{gwilliam_soft_tissue}. 

Contrary to the previous works, Xu et al.~\cite{xu_breast_2013} and Egorov et al.~\cite{egorov_mi_breast} utilized piezoelectric and capacitive sensor arrays for breast examination, respectively. This way, they realised portable and compact devices, an important aspect for BSE and CBE.
Specifically, Egorov et al.~\cite{egorov_mi_breast} employed a three-layer feed-forward neural network for automated lump detection in the silicone tissue model with different location scenarios of various-sized lumps, and compared the results with the performance of manual palpation by naive participants.
Studies showed that the performance increase of the systems compared to the humans as the ball size decreased and depth increased~\cite{gwilliam_soft_tissue, egorov_mi_breast}. 

However, most of these systems were tested using silicone prototypes with primitive structures~(uniform silicone of cubic, cylindrical shapes with a single lump or silicone skin with several lumps) that do not imitate the compliant shape and physical structure of the real breast~\cite{lee_lesion_char, xie_fiber, li_mech_im_tact, gwilliam_soft_tissue, ayyildiz_optoelec, jia_lump_2013, dementyev_mechanical_2021}.

Noting the limitations of the existing systems, glove-based systems are an attractive solution for breast examination assistance with standardised palpation methodology in the natural standing position. 

Glove-based systems consist of sensors sewn or attached to a fabric material and a data acquisition module~\cite{dipietro_glove_survey_08}. They can diagnose by collecting and analysing data from the human hand during manipulation and interaction~\cite{sundaram_learning_2019, buscher_flexible_2015}. Glove-based systems incorporate different sensing modalities that can measure both hand motions~\cite{dong_glove} and forces~\cite{sagisaka_glove, shahrampour_multi_glove} at the contact points~\cite{hammond_toward_2014, liu_glove_based_2017}. Instrumented gloves have been used in different areas such as teleoperation, human-robot interaction~(HRI), master-slave manipulation, hand rehabilitation, and patient monitoring~\cite{force_glove_survey}. 
\begin{figure}[!b]
    \centering
    \includegraphics[width = .85\linewidth, keepaspectratio]{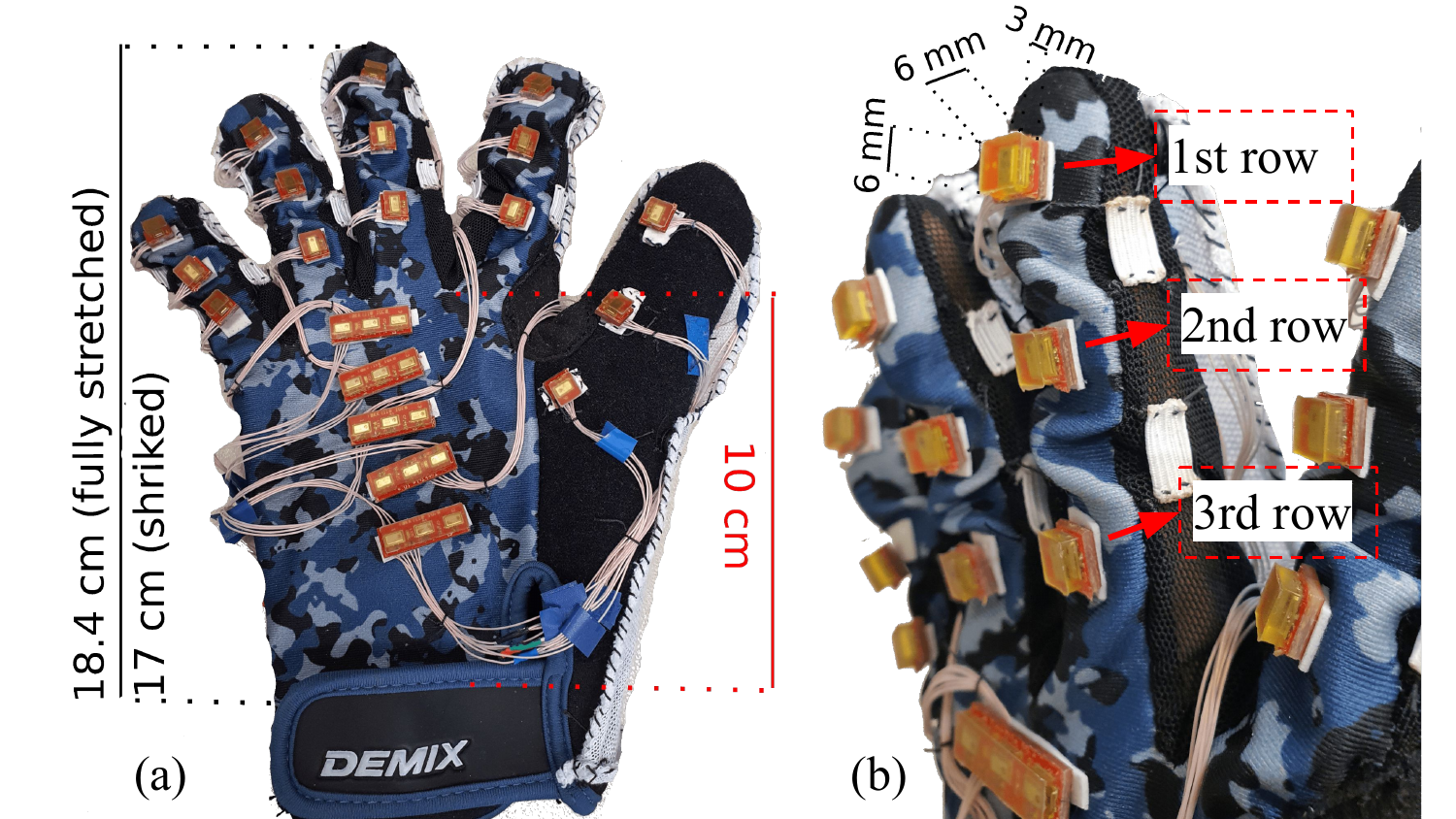}
    \caption{Tactile glove with 30 pressure sensors and two accelerometers (a) top view and (b) side view.}
    \label{fig:glove}
\end{figure}

To the best of the authors' knowledge, a tactile glove was not utilized for a preliminary inspection of lumps in breasts by palpation for assistance and/or co-diagnosis in CBE or self-use in BSE.
To address this gap, we present a tactile glove capable of measuring the force at the points of contact using pressure sensors and processing the acquired data using deep learning~(DL) for breast lump detection and localization (see~Fig.~\ref{fig:breast_with_nodule}). 
The main contributions of our breast lump detection system are: 
\begin{itemize}
    \item the use of a portable tactile glove for the detection and localization of a lump inside silicone breast prototypes;
    \item custom silicone breast prototypes imitating the breast shape with the outer skin, inner substance and chest wall;
    \item a recurrent DL architecture that can predict the breast presence of the lump and determine its size and location;
    \item application of transfer learning on the oncologist-mammologist palpation data for the validation of the CBE with the tactile glove.
\end{itemize}
\begin{figure*}[!t]
    \centering
    \includegraphics[width = .9\textwidth]{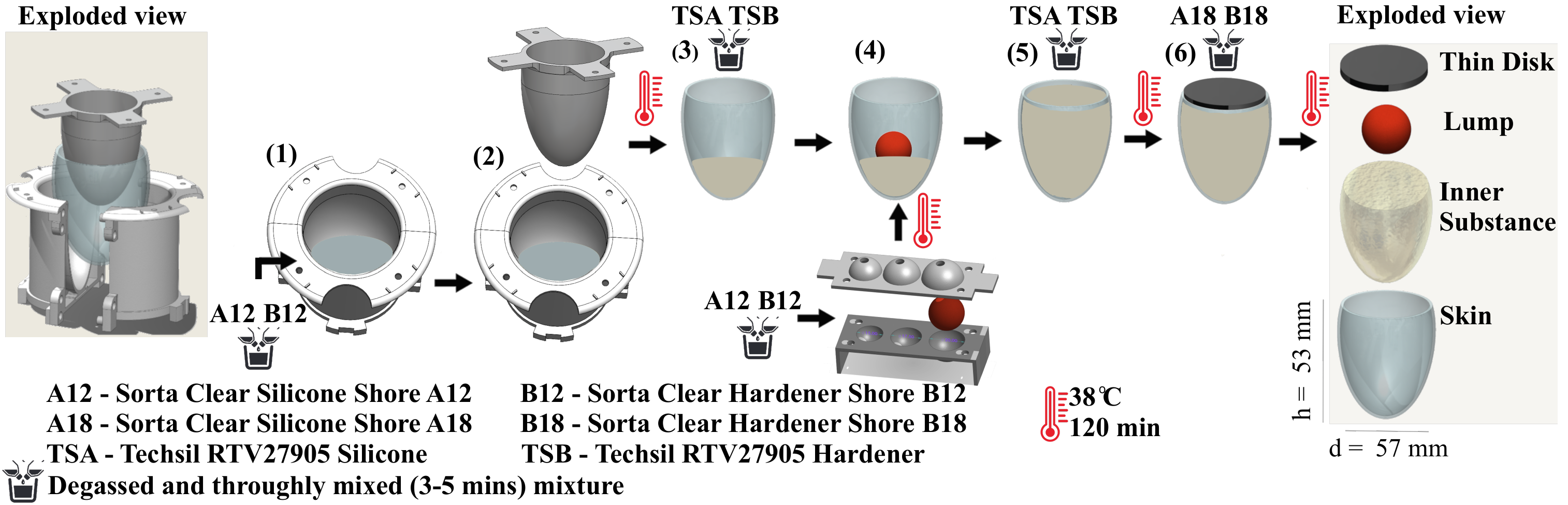}
    \caption{Fabrication pipeline of the silicone breast prototype. It shows the step-by-step SBP casting: (1) and (2) - skin formation by pressing upper mold on lower mold, (3) and (5) - inner substance casting,  (4) - lump formation and placement in between inner substance (skip this step to create a prototype without a lump), and (6) - the creation of the thin disk.}
    \label{fig:sil_production}
\end{figure*}
\section{Hardware}\label{sec:hardware}
In this section, we first describe the implementation of the tactile glove. Then, we present our fabrication procedure for the silicone prototypes mimicking the shape and structure of the breast. We finish the section with a description of our DL architecture for breast lump detection and localization.

\subsection{Tactile Glove}\label{glove_description}
Our fabric-based wearable tactile glove~(see Fig.~\ref{fig:glove}) is capable of measuring pressure distribution and mechanical vibrations at the points of contact. All the wiring is colour-coded, aligned at the back of the hand, and strain-relieved to protect it from wear and tear. The glove is interfaced with a desktop computer (Intel Core i9, 32 GB DDR4 memory, and Linux operating system with real-time Ubuntu kernel) via Robot Operating System~(ROS). 

\subsubsection{Pressure Sensing Module}
The pressure sensing module comprises 30 pressure sensing units distributed over the palm and fingers. Specifically, we are using five Takkstrip 2 tactile sensors (RightHand Labs) that consist of three individual and one $3\times 1$ array of barometers inside vacuum-sealed rubber. The individual units are placed on the distal, intermediate, and proximal phalanges of digits 1-4, the distal and proximal phalanges and the metacarpal bone of the thumb. The data from all 30 sensors is sampled at 100~Hz using an interface board~(TakkFast) with the I2C protocol. ROS-compatible TakkFast board sends the acquired data to the desktop computer using a Universal Serial Bus~(USB). 

\subsubsection{Mechanical Vibration Sensing Module}
The vibration sensing module consists of two MEMS accelerometers (ADXL303, Analog Devices) with a physical bandwidth of 1~kHz, attached to the fingertip of the thumb and index fingers. The data is acquired using a microcontroller board (STM32F3 Discovery, STMicroelectronics) and two custom-made analogue-to-digital converter (ADC) modules (AD7685, Analog Devices) connected in a cascaded way via the serial bus. A single ADC converter module could receive two analog signals from the accelerometer at a time. The microcontroller is connected to the computer through the sound card and samples data at 8 kHz.

\subsection{Silicone Breast Prototypes for Physical Evaluation} \label{hardware:subsec:silicone_prot}
To evaluate the tactile diagnostic glove, we prepared a set of silicone breast prototypes~(SBPs). Specifically, there are two main classes within the SBPs: with and without a lump. The latter has a suspended spherical silicone ball inside, imitating the lump. The spherical balls are of three different sizes (15 mm~(small), 17.5 mm~(medium), 20 mm~(large)) and embedded at three different depths~(shallow/intermediate/deep). Overall, there are 9~\textit{with}~($3$ sizes $\times$ $3$ depths) and 4~\textit{without} a lump identical prototypes~(see Fig.~\ref{fig:set_prototypes_palp}). 
To mimic the structure and shape of the breast, we divide the breast model into three main components: skin, inner substance~(fat tissues) and a thin disk~(to model the chest wall). We experimented with various silicone compounds of different Shore durometers until the oncologist-mammologist stated that the prototypes felt realistic during tactile palpation. 

The fabrication process starts with skin casting using two molds~(upper and lower molds) printed with a selective laser sintering 3D printer (Form 2, Formlabs) using Clear Resin material. Then, we thoroughly mix the silicone parts A and B in 1 to 1 ratio~(SortaClear12, Smooth-On) for three minutes and degas the mixture in a vacuum chamber to remove the air bubbles that can distort the surface and shape of the silicone part. The mixture is poured into the lower mold~(see Fig.~\ref{fig:sil_production}-(1)) and pressed above with the upper mold ensuring even distribution of the material in the gap between the molds ~(see Fig.~\ref{fig:sil_production}-(2)). Both molds are fixed together with screws. We place the mold inside a curing chamber~(Formlabs Form Cure UV Resin, $38^o$C) for two hours to speed up the curing process. The same degassing and curing procedures are applied to fabricate other silicone parts of the breast model.

To level the lump at different depths, we attach the silicone ball to a thin thread and place it inside the dried silicone skin in the lower mold. The silicone balls are prepared by pouring a silicone mixture of parts A and B in 1 to 1 ratio~(SortaClear12, Smooth-On) with a syringe into a 3D-printed mold~(Ultimaker S5) through the small hole at the top~(Fig.~\ref{fig:sil_production}-(4)). The Young's modulus of the silicone ball is 159 $kPa$ which is in the range of the reported values in the literature (25 - 175 $kPa$)~\cite{ayyildiz_optoelec, egorov_mi_breast}. Afterwards, degassed Techsil~(RTV27905) silicone mixture~(1 to 1 ratio) is poured inside the skin to imitate the inner fat of the breast and cured in a curing chamber~(Fig.~\ref{fig:sil_production}-(3) and (5)). Techsil silicone has a slimy and sticky texture even after full curing. Therefore, an encapsulating thin layer of silicone mixture~(SortaClear18, Smooth-on) is poured above to create a ``chest wall'' and cured in a curing chamber~(Fig.~\ref{fig:sil_production}-(6)). The fabrication process details are shown in the supplementary multimedia material accompanying this paper.
In the rest of the paper, lump presence, size and location categories will be defined as follows:
S - small, M - medium, L - large, H1 - top, H2 - middle, H3 - bottom, NL - no lump, WL - with lump.
\section{DL Methods} \label{sec:DL_meth}
We used a number of methods to process the time series data from the pressure sensors to tackle the lump detection and localization problem. 
\subsection{Architectures}
Given nine categorical classes representing prototypes with embedded lumps inside~(see Section~\ref{hardware:subsec:silicone_prot}), their amalgamation into a singular category results in a binary classification scheme: ''with lump" and ''without lump". This consolidation simplifies the DL model selection task via benchmarking and ensures the training of a more robust model. Before training the models, data preprocessing techniques, including normalization and mean imputation, were applied. We rectified missing values and arranged the data into fixed-length tensors. The models that were implemented are listed below. 
\begin{itemize}
    \item \textbf{InceptionTime}: This DL model is designed specifically for classification the time series data. The model includes multiple convolutional filters to capture features at different scales. Its multi-branch structure leads to a better learning of diverse patterns in the data effectively~\cite{ismail_fawaz_inceptiontime_2020}. 
    \item \textbf{XceptionTime}: Inherited from Xception model, this DL method uses depth-wise separable convolutions to reduce the number of tunable parameters.  By separating the spatial features from the channel-wise features during training, this method achieves high performance~\cite{rahiman_xceptiontime}.
    \item \textbf{mWDN (Multi-Weighted Dilated Network)}: This architecture enhances the receptive field without increasing the computational cost by employing dilated convolutions with multiple weighting schemes which capture multi-scale features in the data~\cite{wang_multilevel_2018}. 
    \item \textbf{ResNet}: The Residual Network architecture is designed to mitigate the vanishing gradient problem, i.e. performance plateau, in networks with a large number of layers. Corresponding residual connections were successfully applied in time-series classification tasks~\cite{resnet_ts}.
    \item \textbf{LSTM}: Long Short-Term Memory recurrent neural networks are capable of learning long-term dependencies. LSTMs were designed specifically for sequential data~\cite{hochreiter_long_1997}. 
\end{itemize}
\begin{figure}[!b]
    \centering
    \includegraphics[width = \linewidth]{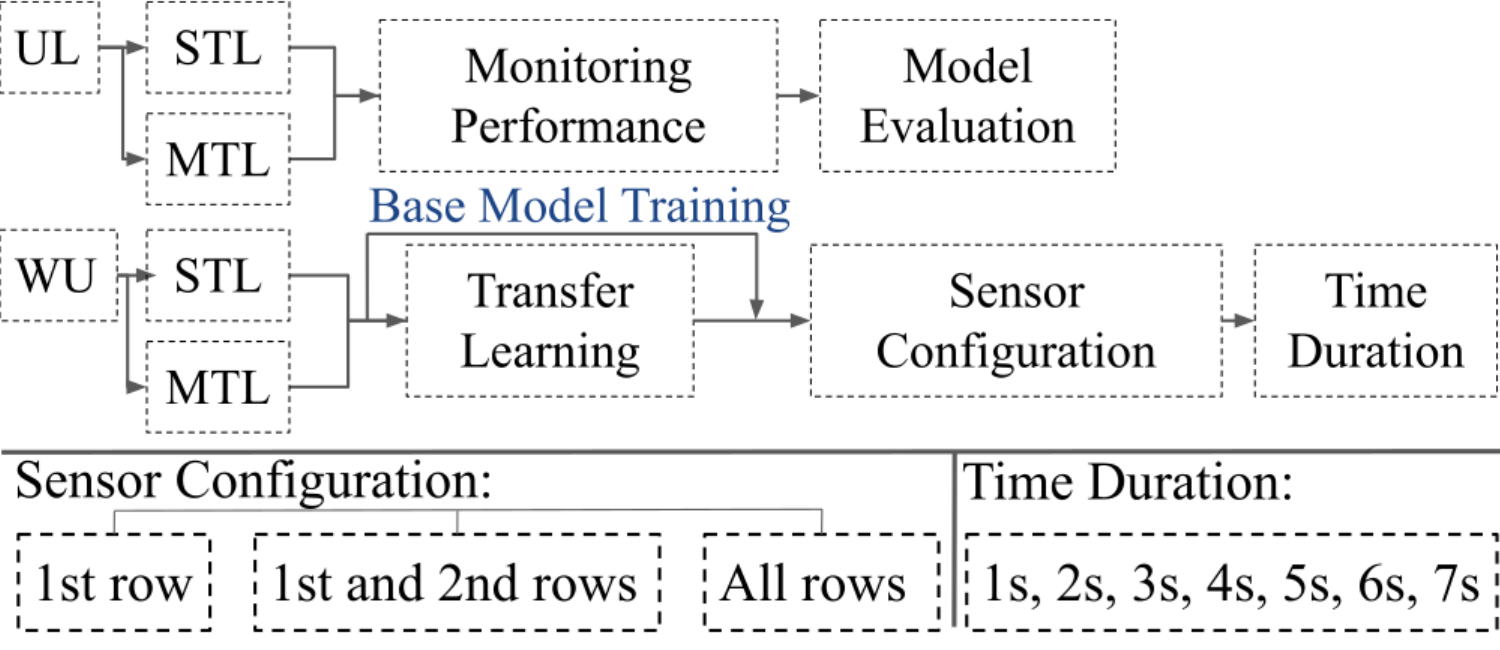}
    \caption{Schematics of the learning approach. UL -- user-level, WU -- within-user, STL -- single-task learning, MTL -- multi-task learning.}
    \label{fig:DL_learning_scematics}
\end{figure}
\begin{figure*}[!t]
    \centering
    \includegraphics[width = \textwidth]{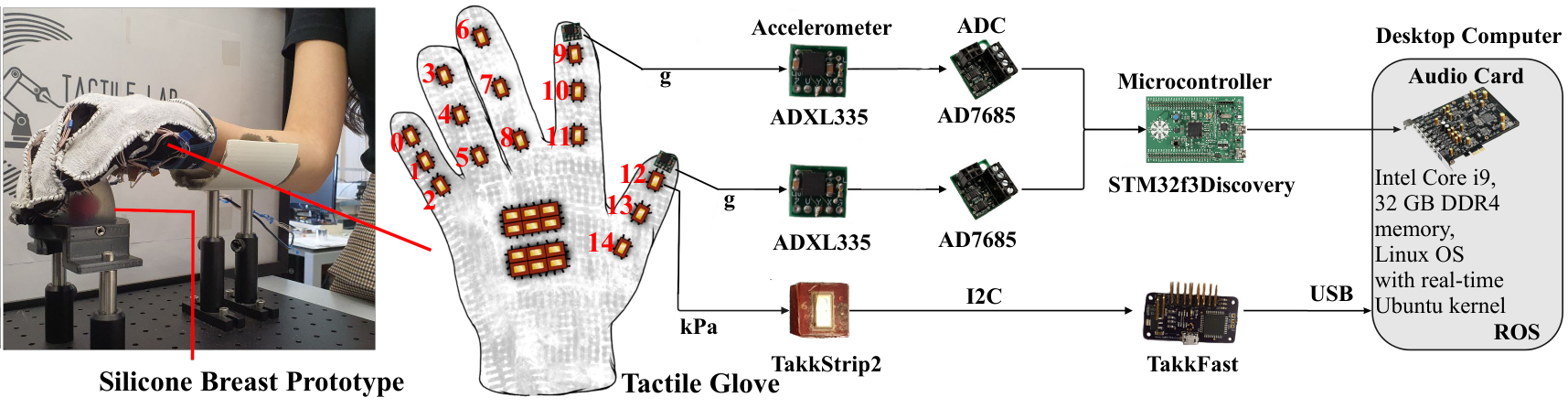}
    \caption{Experimental setup for the lump examination in the silicone breast prototypes. Tactile glove incorporates 3-axes accelerometers (ADXL) and barometers (TakkStrip2), a microcontroller board (STM32) and a  multiplexer (TakkFast).}
    \label{fig:setup_diagram}
\end{figure*}
\subsection{Methodologies}\label{sec:DL_meth:exp_meth}
First, we consider single-task and multitask learning architectures to train the aforementioned models in order to solely detect lumps, or estimate their sizes or locations,  and to perform these tasks altogether. Specifically, \textbf{single-task learning} (STL) involves a single output prediction -- for example, it detects only the presence of lumps --, whereas \textbf{multitask learning} (MTL) predicts simultaneously multiple outputs: presence, size, and position. The latter approach leverages the shared representations and auxiliary information learned from related tasks to improve model performance. Secondly, to assess the generalizability of the base model we apply user-level and within-user data splits. \textbf{User-level} (UL) data split facilitates generalisation assessment across the different users, while \textbf{within-user} (WL) split evaluates the consistency within the individual users. Given that the user-level study may present a lower performance due to the limited number of available data, the inability to capture rare patterns, and the uniqueness of the specific participant's performance, we introduce the \textbf{transfer learning} for the within-user data split approach. We evaluate its adaptability to a new subject and search for the optimal size of the new reams of data for robust performance. Next, we test various sensor configurations and chunks of time duration to find the optimal combination to determine the best setup for accurate and consistent data acquisition in comparable glove-based systems. %

We test combinations of these learning models, data split variations, and the transfer learning process with new participant's data for the following reams of data: within-user data split, various amounts of sensors (configurations), and different time durations~(see Fig.~\ref{fig:DL_learning_scematics}). Thus, we ensure a comprehensive evaluation of the models and their performances under different conditions. We also employed early stopping in all models to avoid overfitting: the training is terminated when the validation loss remains the same for multiple epochs~(patience). 
\subsection{Transfer Learning with Oncologist-Mammologist's Data}
Incremental transfer learning is a machine learning technique where a pre-trained model on one task is adapted to perform a related task. This approach leverages the knowledge gained from the base model training phase and applies it to a new, often smaller, and related dataset. Since the exploratory movements are unique for every person, we consider the transfer of learning between experienced and non-experienced users. Specifically, we enhance the DL performance by training our base models with supplementary data collected from an oncologist-mammologist during the palpation of SBPs. We assume that the oncologist-mammologist is the experienced user, i.e., a professional specialist, ensuring more accurate and reliable results. 
 
It allows us to adapt the model to the new data and evaluate the optimal reams of data required for robust performance on a new user. Below, we describe our pipeline for this process. 
\begin{itemize}
    \item \textbf{Step-by-Step Addition}: The transfer learning process involves the incremental addition of a small number of data points from the oncologist-mammologist's dataset into the training set. This approach replicates a real-world scenario since the initially scarce labelled data increases piece by piece.
    \item \textbf{Balanced Trial Addition}: In each step, when we add one trial from each class of the prototypes with a lump inside~(see Fig.~\ref{fig:set_prototypes_palp}), one trial of the prototype without a lump is added for each trial of the prototype with a lump. Thus, the data remains balanced to prevent the model from becoming biased towards either class. 
    \item \textbf{Monitoring Performance}: After adding each batch of new trials, the model is re-trained, and its performance is monitored using validation loss and accuracy metrics. The process continues until it stops improving the performance, which indicates that the corresponding neural network is sufficiently learned from the new data.
    \item \textbf{Model Evaluation} The model is evaluated on the test set from the oncologist-mammologist's dataset to assess its final performance. This evaluation provides insights on how well the model adapts to the new data.
\end{itemize}
\section{Experimental Procedures} \label{sec:exp_pocedure}
\subsection{Apparatus for Data Collection}
The experimental platform for breast lump detection consisted of the tactile glove and custom-made SBPs~(see Fig.~\ref{fig:setup_diagram}). We collected the data at 160 Hz from the 15 pressure-sensing units and MEMS accelerometers in batches of 1 and 50 measurements, respectively. Given that we have the sensors located on the fingers and palm, the data collection was performed using the sensors located on the fingers only: based on the consultation with oncologist-mammologists and research on clinical breast examination protocol~\cite{henderson_breast_2024}, it was decided to perform palpation using only the fingers. Likewise, the duration of every palpation was set to 7~s. 
\subsection{Data Collection}

We conducted data collection experiments with ten healthy participants~(5 females~(F) and 5 males~(M) with a mean age of 22 (SD = 2)) from the university community. The study protocol was approved by the Institutional Research Ethics Committee~(IREC) of Nazarbayev University and written informed consent was obtained from all participants.

\begin{figure}[!t]
    \centering
    \includegraphics[width = .88\linewidth]{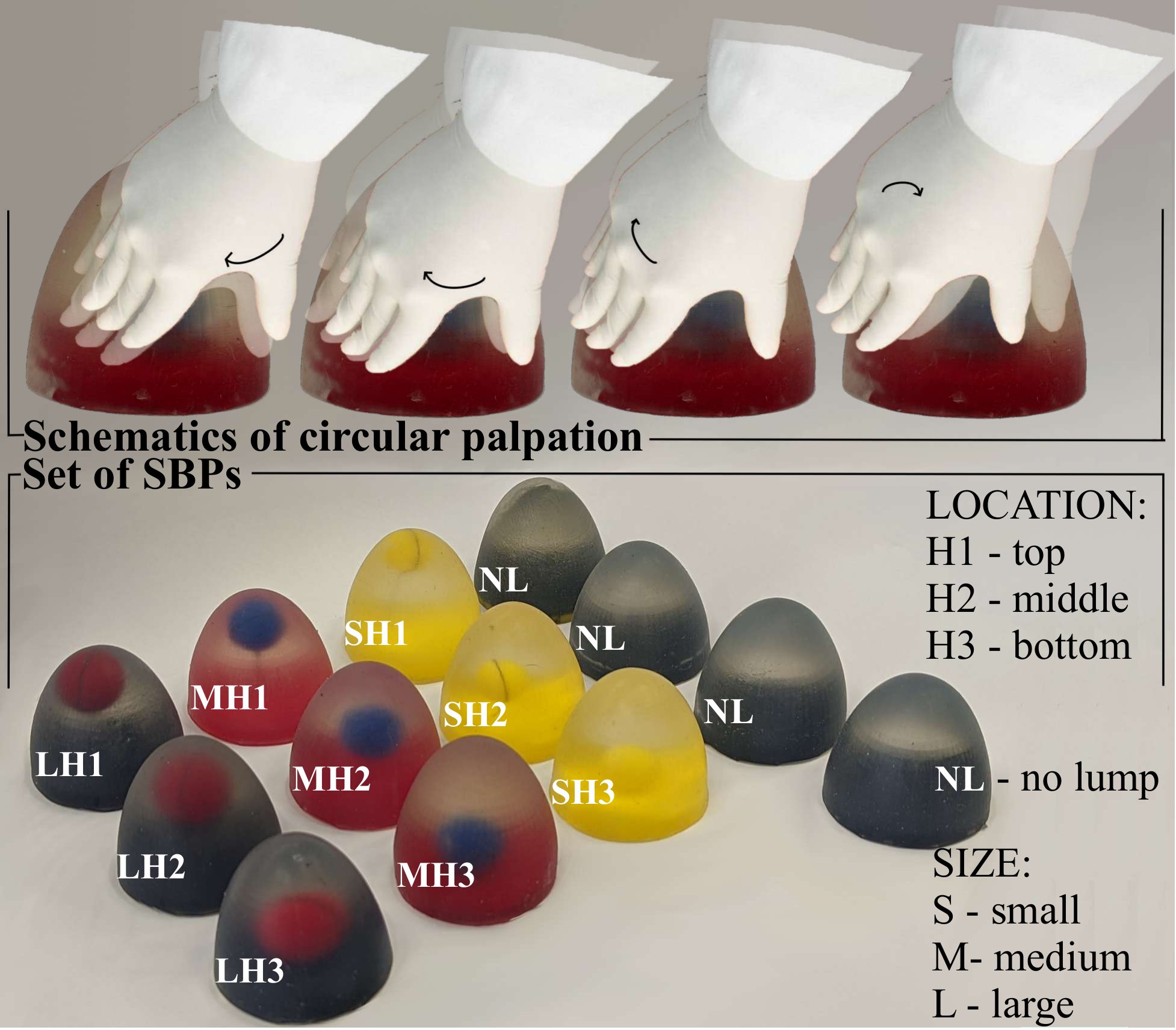}
    \caption{Data collection. (Top) Schematic illustration of the silicone breast prototype (SBP) examination with a circular motion by a human participant; (Bottom) The set of SBPs.}
    \label{fig:set_prototypes_palp}
\end{figure}

During the experimental sessions, the participants were provided with a video demonstration of a doctor palpating the SBP by applying circular motions using only fingers for 7~s~(see~Fig.~\ref{fig:set_prototypes_palp}). The data collection for each trial started when contact between the object and fingers occurred. The contact was detected by thresholding the variation of the total pressure applied by the subject. The total pressure was estimated by summing the derivatives of the output values of every pressure sensor on the glove. We assumed that the contact occurred when the absolute value of this sum exceeded a threshold $c_2$ determined experimentally. 

The participants were asked to palpate the 9 prototypes with a lump 32 times each, ensuring all fingertips were in contact with the object. To balance the data of prototypes \textit{with} ~(9 pieces) and \textit{without} ~(4 pieces) a lump, each of the latter ones was palpated 72 times. On each trial, we randomly presented one of the 13 prototypes to the participants.
This resulted in 576 trials for each subject~(9 prototypes \textit{with} a lump $\times$ 32 palpations $+$ 4 prototypes \textit{without} a lump $\times$ 72 palpations). Overall, the experimental procedure was divided into two sessions with small breaks in between to stretch the hand and relax; it took about 3 hours to complete.

In addition to the naive participants, we invited a mammologist from the municipal oncological prophylactic centre to conduct data collection experiments. The same experimental protocol was applied in the data collection procedure. The number of trials was twice as small as that of the naive participants, resulting in 288 trials~(9 prototypes \textit{with} a lump $\times$ 16 palpations $+$ 4 prototypes \textit{without} a lump $\times$ 36 palpations). In the rest of the paper, the following dataset will be referred to as oncologist-mammologist data.
\subsection{Data Preprocessing}
Each experimental trial (one participant palpating one prototype) generated a 1$\times$15  array of pressure measurements during $7$~s at $160$~Hz, which resulted in  1120 samples; we also recorded a 1$\times$4 array of accelerometer measurements, which was not used in this study (56,000~samples from two 2-axes accelerometers during 7~s at $2$~kHz).
For the pressure sensing unit data, empty data points were filled using mean imputation, and the entire dataset was normalised to ensure consistent scaling across all features. Similarly, the accelerometer data underwent preprocessing steps, wherein missing data points were addressed through mean imputation, and the entire dataset was normalised to maintain uniformity in feature scaling.
The data was transformed in the two-dimensional array where each consequent 15 rows represents the data of 15 distinct pressure sensing units~(referred to as ``feature'') for each experimental trial.

The dataset was split into train, validation, and test sets using two different stratification approaches described in Section~\ref{sec:DL_meth:exp_meth}. 
\begin{itemize}
    \item \textbf{User-level data split}: 
    We randomly allocated data from: 7 participants for the training set, 2 participants for the validation set, and 1 participant for the test set. This resulted in 4032~(9~prototypes \textit{with} lump~$\times$~32~trials~$\times$~7~participants and 4~prototypes \textit{without} lump~$\times$~72~trials~$\times$~8~participant), 1152~(9~prototypes \textit{with} lump~$\times$~32~trials~$\times$~2~participant \textit{with} lump and 4~prototypes \textit{without} lump~$\times$~72~trials~$\times$~1~participant \textit{without} lump) and 576 trials in train, validation and test sets. This ensured that no individual appeared in more than one of the sets. In total, there were 1120$\times$576 samples per participant. 
    \item \textbf{Within-user data split}: 
    From each participant, we selected a proportional number of trials \textit{with} and \textit{without} lumps in a stratified approach for the training, validation, and test sets. Specifically, we ensured that data from the same individuals were distributed across all three sets, maintaining an equal representation of trials \textit{with} and \textit{without} lumps. 
    From each of the 10 participants, we allocated 360~(20~trials~$\times$~9~prototypes \textit{with} lump and 54~trials~$\times$~4~prototypes), 144~(4~trials~$\times$~9~prototypes \textit{with} lump and 9~trials~$\times$~4~prototypes) and 72~(4~trials~$\times$~9~prototypes \textit{with} lump and 9~trials~$\times$~4~prototypes) trails for the train, validation and test sets, respectively. The same stratification was applied for the oncologist-mammologist's data~(24~trials~$\times$~9~prototypes \textit{with} lump and 54~trials~$\times$~4~prototypes).
\end{itemize}
\begin{table*}[t!]
    \caption{Performance of different DL methods for lump detection classification using pressure sensor data. mWDN - Multilevel Wavelet Decomposition Network, ResNet - Residual Neural Network, ResCNN - Resilient Convolutional Neural Networks, LSTM - Long short-term memory, FCN - Fully Convolutional Network, BiLSTM - bidirectional LSTM}
    \centering
    \begin{tabular}{cccccccccccc}
    \toprule
    Models & InceptionTime & XceptionTime & XResNet1d34 & mWDN & ResNet & ResCNN & LSTM-FCN & FCN & LSTM & BiLSTM \\ 
    \midrule
    Accuracy & \textbf{94.4\%} & 94.3\% & 92\% & 92\% & 91.1\% & 90.6\% & 89.8\% & 88\% & 83.9\% & 83.5\% \\ 
    \bottomrule
    \end{tabular}
    \label{tab:benchmarking}
\end{table*}
\subsection{DL Model Training }
In this section, we describe the training procedures and evaluation metrics used to assess the performance of various models under different conditions. We employed the models (Sec.~\ref{sec:DL_meth:exp_meth}) with different hyperparameters and derived a benchmark table to identify the best-performing one, which is most apt for the binary classification problem with the provided data. Given the computational limitation, the benchmarking approach saves significant computational resources and time. Additionally, benchmarking offers a standardised dataset evaluation ensuring a fair and consistent comparison of pre-evaluated results. 
\begin{figure}[!b]
    \centering
    \includegraphics[width=.8\linewidth]{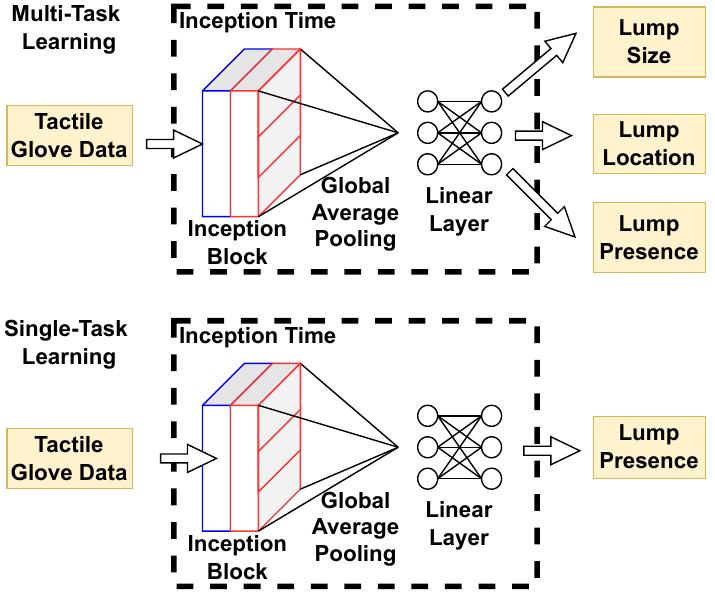}
    \label{fig:DL_single}
    \caption{Architectures of multi-task and single-task models.}
    \label{fig:DL_single_multi}
\end{figure}
\subsubsection{Benchmarking Best Model}
After evaluating various models, listed in Section~\ref{sec:DL_meth}, the InceptionTime was identified as the best-performing model for our binary classification problem. The hyperparameters for the InceptionTime model were optimised using a grid search approach, with the final model trained using a learning rate of 0.001, batch size of 32, and 10 epochs. We trained single task and multitask models based on InceptionTime~(see Fig.~\ref{fig:DL_single_multi}).
\subsubsection{Single Task -- User-level and Within-User}
A single-task model was trained on the binary classification problem of lump presence detection with user-level and within-user data split, separately. The InceptionTime model was employed, trained with a learning rate of 0.001, batch size of 32, and 100 epochs. The user-level split ensured that data from different users were strictly separated into training, validation, and test sets. The within-user split included data from all users in each of the training, validation, and test sets, ensuring a proportional representation of trials with and without lumps across the different sets.
\subsubsection{Multitask -- User-level and Within-User}
The multitask model was trained on the binary classification problem of lump presence detection, as well as on lump size and position classification, with user-level and within-user data splits, separately. The InceptionTime model was employed, trained with a learning rate of 0.001, batch size of 32, and 100 epochs. The model outputs three predictions: lump presence, lump size, and lump position.

\subsubsection{Multitask -- Within-User -- Transfer Learning with oncologist-mammologist's Data}\label{sec:exp_platform:sys-config}
We further applied transfer learning on the multitask model trained with a within-user dataset split with additional oncologist-mammologist's data. The InceptionTime model, pre-trained on the dataset collected from 10 participants~(further referred to as the ''initial dataset"), was trained with a learning rate of 0.0001 and a batch size of 16. The incremental transfer learning process involved gradually adding one trial with a lump and one trial without a lump from the oncologist-mammologist's data, monitoring the model's performance to prevent overfitting.

\subsubsection{Multitask -- Within-User -- Transfer Learning with Oncologist-Mammologist' Data with Different Duration and Sensor Configurations}
We run experiments to find an optimal sensor placement and data acquisition duration for the lump detection and localization task via a glove-based system. This comprehensive analysis aims to determine the most efficient setup for accurate and consistent data capture of comparable glove-based systems. 
The sensors were labelled from 0-14 as shown in Fig.~\ref{fig:setup_diagram}.
\begin{itemize}
    \item \textbf{Sensor Configurations}:
    \begin{itemize}
        \item \textbf{1st row}: Labels: 0, 3, 6, 9 and 12. 
        \item \textbf{1st and 2nd row}: Labels: 0, 1, 3, 4, 6, 7, 9, 10, 12 and 13.
        \item \textbf{All}:  Labels: from 0 to 14.
    \end{itemize}
    \item \textbf{Time Duration}: The total length of the data 
    duration ranging from 1 to 7~s was used to evaluate the model's performance across different time frames, assessing the trade-off between data length and model performance.
\end{itemize}

This experiment extended the transfer learning process of the model described above with various sensor configurations and data duration to determine the most effective setup for clinical applications of glove-based systems. This comprehensive analysis aims to determine the most efficient setup for accurate and consistent data capture of comparable glove-based systems and to ensure robustness and applicability in real-world clinical settings.

\begin{table*}[!t]
\centering
\caption{Lump Detection Accuracy for Single-task/Multitask Model with Within-User~(WU) and User-Level~(UL) Data Splits and Transfer learning.}
\begin{tabular}{c c c c c c}
\toprule
\multirow{3}{*}{\textbf{Data Split}}  & \multirow{3}{*}{\textbf{Training Phase}} & \textbf{Single-task} &\multicolumn{3}{c}{\textbf{Multitask}}\\ \cmidrule(lr){3-3}
\cmidrule(lr){4-6}
 &  & \textbf{Lump Presence } & \textbf{Lump Presence} & \textbf{Size} & \textbf{Position}\\ 
 &  & \textbf{Accuracy~(\%)} & \textbf{Accuracy~(\%)} & \textbf{Accuracy~(\%)} & \textbf{Accuracy~(\%)}\\ 
 \midrule
WU  & Base Model Training & 48.61 & 46.01 & 21.70 & 29.86\\  \midrule
\multirow{10}{*}{UL} & Base Model Training & 81.52 & 82.22 & 67.08 & 62.63\\
    & Testing on oncologist-mammologist's without transfer learning & 51.38 & 51.73 & 20.48 & 19.79\\
                & Transfer learning with 1 trial  & 77.78 & 51.73 & 18.40 & 19.44 \\
               & Transfer learning with 2 trials & 67.01 & 51.73 & 19.79 & 23.95 \\
               & Transfer learning with 3 trials & 80.20 & 52.08 & 24.65 & 20.13\\
               & Transfer learning with 5 trials & 93.05 & 66.67 & 47.91 & 45.13\\
               & Transfer learning with 6 trials & 93.05 & 77.43 & 65.27 & 60.76\\
               & Transfer learning with 7 trials & 92.36 & 86.11 & 70.48 & 66.31\\
               & Transfer learning with 8 trials & 87.50 & 91.66 & 74.30 & 74.31\\
               & Transfer learning with 9 trials & \textbf{95.48} & 85.06 & 73.26 & 67.36 \\
               & Transfer learning with 10 trials& 93.05 & 92.01 & 82.29 & 74.30 \\
               & Transfer learning with 11 trials& 94.45 & 93.05 & 80.55 & 72.91 \\    
               & Transfer learning with 12 trials& 93.51 & 92.70 & 79.86 & 79.86 \\    
               & Transfer learning with 13 trials& 94.47 & 87.75  & 80.55 & 74.30 \\ 
               & Transfer learning with 14 trials& 95.43 & 92.01 & 86.80 & 65.62 \\ 
               & Transfer learning with 15 trials& 94.47 & \textbf{95.01} & \textbf{88.54} & \textbf{82.98} \\ 
\bottomrule
\end{tabular}
\label{tab:initial_fine_tuning}
\end{table*}
All experiments were conducted using an Nvidia DGX-2 system, which features 8 Nvidia Tesla V100 GPUs with a total of 256 GB GPU memory, dual 20-core Intel Xeon E5-2698 v4 processors at 2.2 GHz, and 512 GB of RAM.
\subsection{Human Perception Assessment}
We conducted manual palpation experiments on SBPs with ten additional healthy participants~(5~M and 5~F with the mean age of 22 (SD = 2))) from the university community with an average age of $25$. This experiment aims to compare the performance of the naive participants in lump detection and localization tasks with the performance of our system. The participants were given time to familiarize themselves with the prototypes. Given that the outer skin of the silicone prototypes is transparent, the participants were blindfolded during the entire experimental procedure to eliminate any visual clues. Then, the participants were asked to palpate the prototypes with bare hands in circular motions~(see~Fig.~\ref{fig:set_prototypes_palp}) using their fingers for 15-20~s and report if there was a lump. We presented the prototypes in random order resulting in 144 trials for each subject (8 trials $\times$ 9 \textit{with} a lump  + $18$ trials $\times 4$ \textit{without} a lump). Overall, the experimental procedure took about 1.5 hours to complete with small breaks to avoid fatigue. 
\section{Results}\label{sec:results}
\begin{figure*}[!t]        
    \centering
    \includegraphics[width =.85 \textwidth]{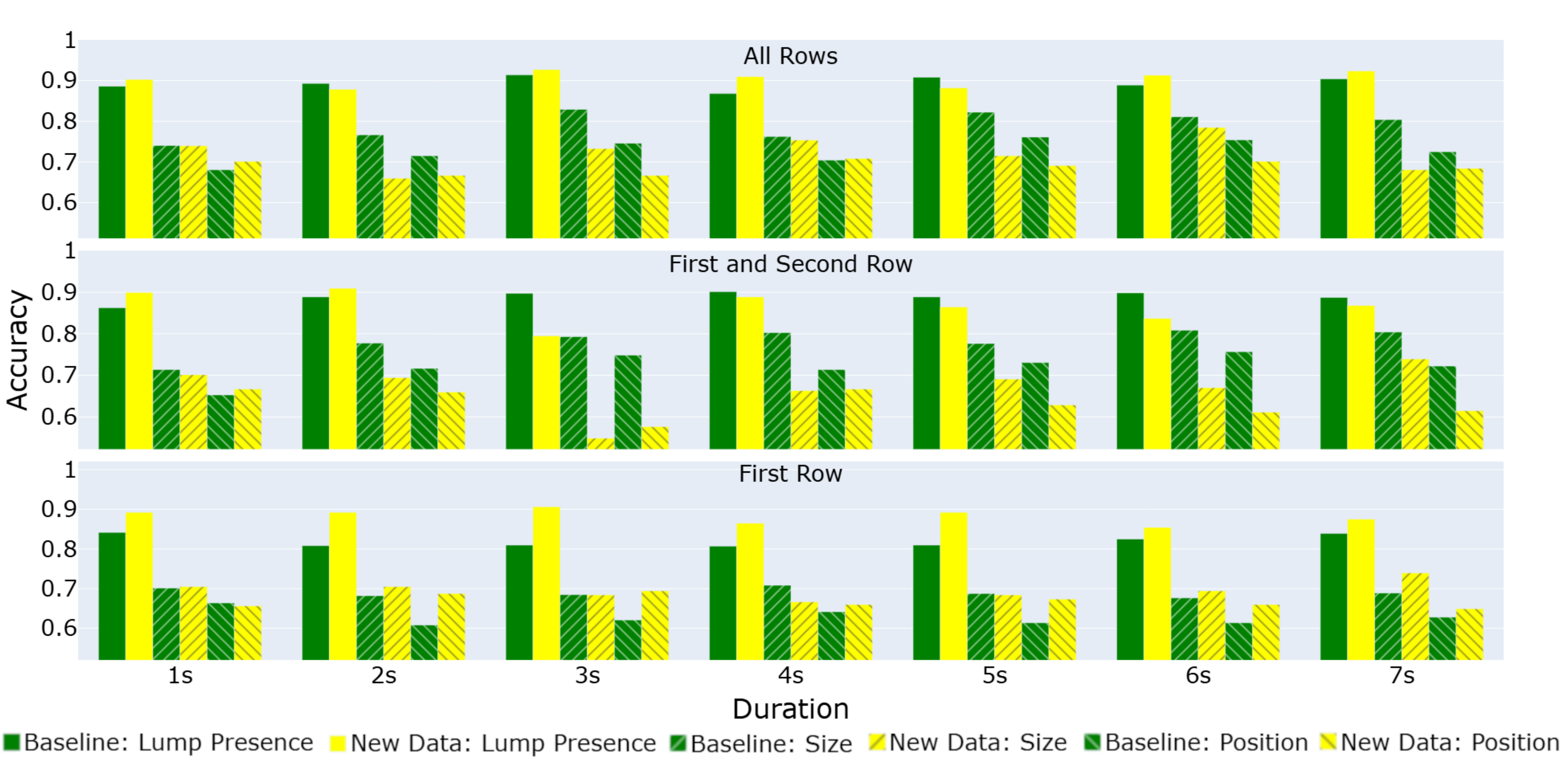}
    \caption{Model's performance on the variation of sensor configuration~(5, 10, or 15 sensors for first, first and second, or all rows, respectively) and data time durations~(160 samples per second).}
    \label{fig:all_bar}
\end{figure*}
\begin{figure}[!t]
        \includegraphics[width=0.85\linewidth]{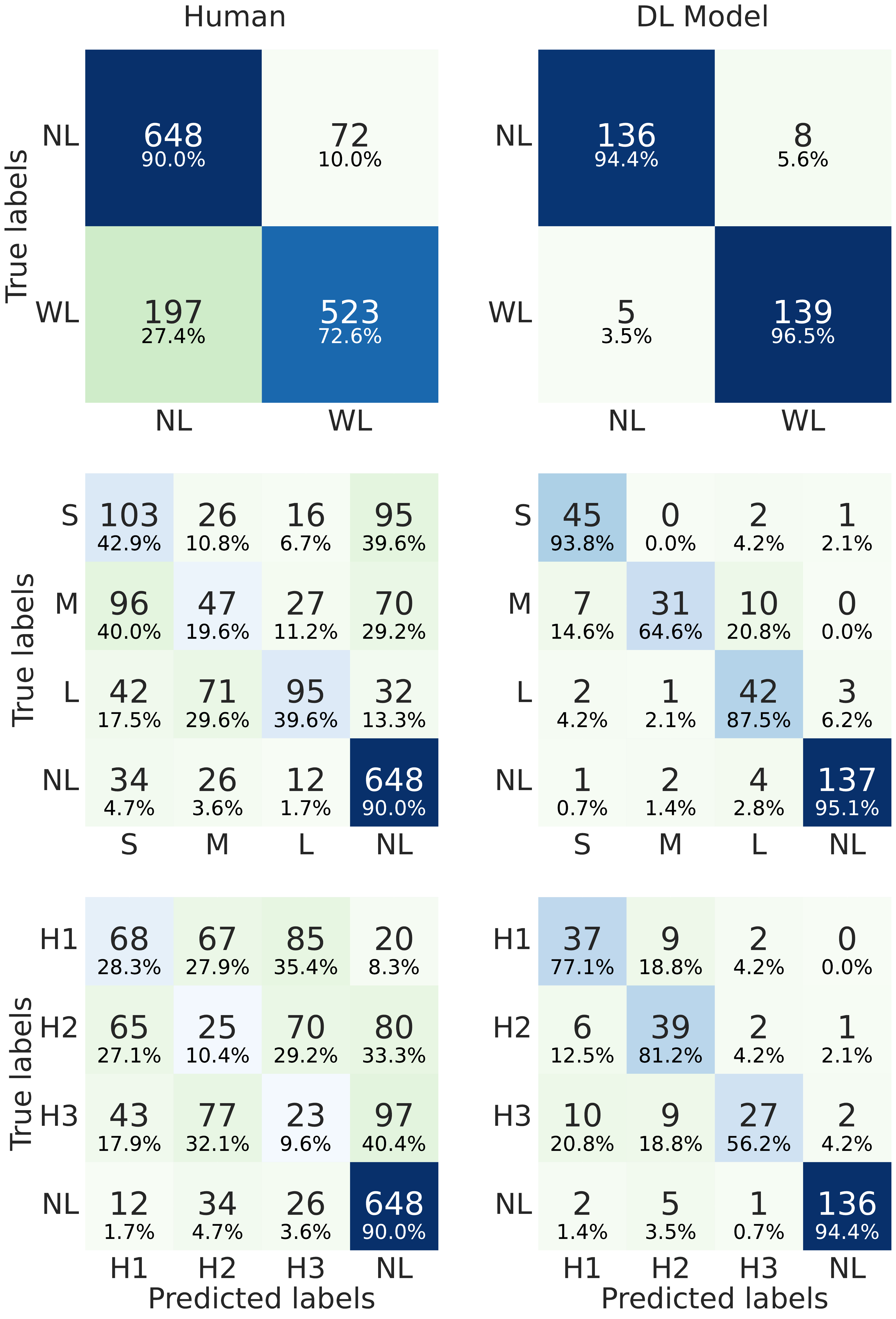}
        \caption{Confusion matrices of the lump presence, size, and location classifications for human and DL model}
        \label{fig:confusion}
\end{figure}  
\subsection{Lump Detection with DL Methods}
 Among the ten architectures we tested for the benchmarking~(see Table~\ref{tab:benchmarking}), XcetionTime and InceptionTime models showed the best results, with an overall test accuracy of 94.4\% and 94.3\%, respectively. Meanwhile, LSTM and BiLSTM, classic methods for dealing with sequential data, exhibited lower performance with an accuracy of 83.9\% and 83.5\%, respectively. We chose the InceptionTime as the best performing model~(highlighted in bold in Table~\ref{tab:benchmarking}) due to a smaller validation error than XcetionTime. Additionally, we trained the InceptionTime model on the data from two accelerometers. The result of the lump presence binary classification task was 60.6\% that was lower than for the pressure sensors and, therefore, disregarded in this study. 

The single-task and multitask models, trained on data with user-level split, achieved an accuracy of 48.61\% and 46.01\%, respectively, in a binary classification task to detect the lump presence. However, the same models trained on the data with within-user split showed superior performance with an overall accuracy of 81.52\% and 85.27\%, correspondingly. The initial accuracy~(before transfer learning) of the single-task model was 81.52\%, which dropped to 51.38\% when tested on the oncologist-mammologist's data~(see Table~\ref{tab:initial_fine_tuning}). However, during the transfer learning process, the accuracy gradually improved to 95.48\% on the ninth step of the incremental addition of oncologist-mammologist's data, after which, no further improvements were observed. 

As for the multitask model, the initial accuracy was 88.05\%, 70.69\% and 63.47\% for the binary classification of the lump presence, and multiclass classification of lump size and location, respectively. Like in the case of a single-task model, the test of \textit{base model} training on the oncologist-mammologist's data resulted in an accuracy of 50\%, 11.11\% and 11.45\% in the lump presence, size and location classification, respectively. Therefore, we again applied incremental transfer learning to boost the model's performance on the new user. It required 15 steps of new data addition to achieve the accuracy of 95.01\%, 88.54\% and 82.98\% in the lump presence, size and location classification, correspondingly~(see Table~\ref{tab:initial_fine_tuning}). The results on different sensor configurations and time durations (see Fig.~\ref{fig:all_bar}) showed that the increase of the number of rows, i.e. number of sensors,  and time duration are followed by the increase in accuracy for the binary classification task: the lump presence.
\subsection{Human Perception Study}
The confusion matrices shown in Fig.~\ref{fig:confusion} offer a comprehensive overview of the response patterns, contributing to our understanding of how object perception varies between experiment conditions.

The overall classification accuracy is $81,3\%$ with the True Positive~(TP) and the False Positive~(FP) rates of $72.6\%$ and $10\%$, respectively. Additionally, we recorded the participants' responses on the estimated size and location of the perceived lump. Figure~\ref{fig:confusion} shows that the rate of False Negative~(FN) surges as the size of the lump diminishes. The FN rate is reported to be 39.6\%, 29.2\% and 13.3\% for the small, medium and large sizes of the lump, correspondingly. 
The same pattern is noticeable in the location variation of the lump. The percentage of the FN responses drastically increases to 33.3\% and 40.4\% when the lump is placed at the medium and deepest location inside the prototype. However, the FP percentage of the responses remains within the range of 1.7\%-4.7\% for all the cases. Interestingly, the false negative diagnosis rate increased when the lump was placed deeper and the lump size was decreased.
\section{Discussion}\label{sec:discussion}
This work is the first study to examine the presence of the lump using active circular palpation similar to the medical examination of SBPs via Tactile Glove using DL. In a series of experiments, this study examined the performance of the human participants and the DL model in perceiving the lump and discriminating its size and location.  
The DL results were compared to the manual palpation of ten healthy participants with no medical degree. In our study, SBPs contained stiff inclusions of various sizes and locations.  Owing to numerous factors, the properties of contacts with these SBPs are difficult to predict. However, successful palpation of the SBPs, depends on the sense of touch. In this section, we discuss the obtained results, which are inherently inter-winded with human perception, especially with the sense of touch. 

\subsection{Human Perception vs DL model} 
The comparative analysis of human and machine learning performances reveals several insights. Both of them have similarly low FP rates, which is not a critical scenario because the patient would undergo additional examination, resulting in minimal stress. However, the FN output can result in a missed, possibly cancerous, lump. The results suggested that the performance of the DL model is superior to the human ability to detect the tumour-like inclusions embedded in tissue-like soft silicone samples~(see Fig.~\ref{fig:confusion}). We can see a distinct classification pattern for the size and location predictions by the DL model. In the case of the manual palpation by human participants, they could not discriminate the lump size and location accurately. A similar result was reported in  \cite{ayyildiz_optoelec}: the DL model achieved 24\% higher accuracy than human subjects in binary classification task (96.5\% versus 72.6\%). Moreover, the  reported  TP rate ($\sim 67$ \%) for medium-sized lump detection in human psychophysics experiments is the same as in our experiments.

\subsection{Tactile Glove and Silicone Breast Prototypes}
In the real case scenarios, various contact pressures within the breast are expected: submuscular lump placement will have contact pressures due to pectoralis major muscle contractions; posterior pressures of the posterior lump surface; and body force pressure supporting the gravitational load of the lump along the base of it~\cite{Implants_mechanics_pressure_formula_2020}. Thus, our current DL-based tactile imaging approach may not achieve the same results on real breasts.  However, we observed an advantageous trend in lump detection using the glove~(DL Model) since FNs are smaller than in lump detection without the use of the glove~(Human).

The normal breast tissue is an inhomogeneous
structure containing different tissue layers, including fat, glandular tissue and skin. Owing to the highly non-linear, viscoelastic and anisotropic properties of these layers, the properties of contact are difficult to predict. Moreover,  these properties vary with
age, hydration, physical condition and disease.  As shown in breast mechanics studies, the skin, which has 1 to 3 mm in thickness, changes its properties over time~\cite{Mechanics_breast_2007}. The thickness of the skin layer in all our breast prototypes was 2~mm. We did not vary this and other parameters intentionally,  including the stiffness of the lumps themselves and the softness of the internal silicone gel wrapped by the skin layer. 

\subsection{User-Level Data Split vs Transfer Learning with Within-User Data Split}
Given that we used user-level and within-user data splits, the results indicated considerably poor performance~(accuracy is lower than the probability of a pure guess) with the user-level data split. This means moderate performance when the model is evaluated on entirely unseen individuals, highlighting the challenge of generalising across different users~\cite{ROSTER2022112306}.
On the other hand, models trained on the within-user data split, in comparison, showed superior performance in both single-task and multitask learning approaches. 

However, the within-user data split provides us with a model that cannot be generalised to a new user of the glove-based system. It can be seen when the model is tested on the oncologist-mammologist's data. Therefore, to address the problem of generalizability we introduced a transfer learning approach to our analysis as it can help significantly improve the performance on a new data~\cite{Ismail_Fawaz_2018}. This is particularly useful in clinical settings where collecting labelled data is challenging and time-consuming. Applying transfer learning to the model with data from different sources~(in this case, data collected by doctors) improves its ability to generalise to new and diverse scenarios. This enhances the model's robustness and reliability when deployed in real-world applications.

Interestingly, during training, single-task learning ({19 epochs}) converged faster than multitask learning ({43 epochs}). Presumably,single-task learning focuses on a specific classification task, whereas multitask learning deals with multiple tasks concurrently. On the other hand, the multitask approach appears to be computationally efficient for deployment since it works on several tasks simultaneously as noted also in~\cite{Samala_2017}.

We used the InceptionTime DL model in the aforementioned classification approaches. The superior performance with respect to other models could be its stacked architecture consisting of ensemble filter bank modules with weights trained via backpropagation. The model extracts latent hierarchical features of different resolutions using filters with various lengths~\cite{ismail_fawaz_inceptiontime_2020}.

\subsection{Effect  of Sensory Homunculus}
Usually, the more data we feed into a DL model, the more accurately we can approximate the multidimensional function between the input and output. Our results do not follow this paradigm in the following particular case: when transfer learning is applied to our model for the whole time interval, the accuracy on size and location of the lump detection is higher for the First row (Distal phalanges) than for the combined First and Second row (Distal and Middle phalanges) as illustrated by dashed yellow bars in Fig.~\ref{fig:all_bar}. The recognition rates in size and position were 76\% and 66\% versus 72\% and 62\%, respectively. We believe that the reason for such controversial results is hidden in the exploratory movements of the particular user on whom, the DL model was transfer learned. The user could rely on fingertips more than on other sides of the fingers. Indeed, the density of mechanoreceptors in tactile sensory innervation of the hand is different for fingertips, phalanges and palm; there are more mechanoreceptors on the Distal phalanges~\cite{johansson2009coding}. 

Our other observation seen in Fig.~\ref{fig:all_bar} is that the optimal sensor configuration and the time duration needed to detect lump presence are the following. One second was enough to finish at least one circular motion (Fig.\ref{fig:set_prototypes_palp}). Only the First and Second rows are enough to achieve 90\% in recognition rate, which is equal to the result for All rows. This result complies with the breast palpation protocol, in which only distal (First raw) and middle (Second row) phalanges are used for exploratory movements~\cite{henderson_breast_2024}.
\section{Conclusion}\label{sec:conclusion}
In this work, we showed how the wearable TI technology, i.e., tactile glove and DL methods, can augment tactile sensitivity and enable the collection of diagnostic information for breast lump detection. Potentially, this technology will enable healthcare providers, including nurses, to detect lumps without the need for specialized imaging equipment or exposing patients to radiation, thus, enabling frequent intermediate checks. These goals align with the ones stated in worldwide healthcare initiatives aiming to reduce the burden on specialised diagnostic services~\cite{WHO_initiative_2023}.
\section*{Acknowledgements}
We thank MSHE Kazakhstan Grant number AP26199897.
J. A. Corrales Ram\'on was funded by the Spanish Ministry of Universities through a ’Beatriz Galindo’ fellowship (Ref. BG20/00143), by the Spanish Ministry of Science and Innovation through the research project PID2023-153341OB-I00, by the Interreg VI-B SUDOE Programme through the research project ROBOTA-SUDOE (Ref. S1/1.1/P0125) and by the European Union (European Regional Development Fund - ERDF).
\balance
\bibliographystyle{IEEEbib}
\bibliography{arxiv_template}
\end{document}